\documentclass[aps,preprint]{revtex4}
\usepackage{amsfonts}
\usepackage{amsmath}
\usepackage{amssymb}
\usepackage{graphicx}
\usepackage{subfigure}
\usepackage{epstopdf}
\usepackage{float}
\begin{document}
\preprint{CTP-SCU/2020018}
\title{Universal thermodynamic extremality relations for charged AdS black hole surrounded by quintessence}
\author{Qiuhong Chen$^{a}$}
\email{2017141221016@stu.scu.edu.cn}
\author{Wei Hong$^{a}$}
\email{thphysics_weihong@stu.scu.edu.cn}
\author{Jun Tao$^{a}$}
\email{taojun@scu.edu.cn}
\affiliation{$^{a}$Center for Theoretical Physics, College of Physics, Sichuan University, Chengdu, 610065, China}
	
\begin{abstract}
The perturbation correction of general relativity has obtained the general relationship between entropy and extremality. The relation was found to exactly hold for the charged AdS black hole, and extended to the rotating and massive gravity black holes. In this paper, We confirm the universal relation for the four dimensional RN-AdS black hole surrounded by the quintessence, and find a new extremality relation between black hole mass and the normalization factor $a$ related to the density of quintessence. Besides, we also find that when the black hole entropy increases, the correction constant is negative, the black hole mass and the mass-charge ratio decreases, hence this black hole displays WGC-like behavior. Moreover, we derive the general relationship on nonlinear electrodynamic source couples to AdS spacetime with quintessence field. We find another new extremality relation between black hole mass and the magnetic monopole charge $\beta$. 
\end{abstract}
\maketitle\tableofcontents

\section{Introduction}
Quantum gravity combines the principles of general relativity and quantum theory, which expected to be able to provide a satisfactory description of the microstructure of spacetime on the Planck scale. This scale is so far from current experimental capabilities that it is almost impossible to empirically test the theory of quantum gravity along the standard route. Quantum gravity leads to the area dependence of these solutions, that cannot be clearly explained in terms of the original Einstein gravity \cite{Strominger:1996sh}. Many aspects of quantum gravity can be analyzed within the framework of efficient field theory. String theory is widely regarded as the complete description of quantum gravity. This theory is also thought to acknowledge an astronomical vacuum that behaves as the efficient field theory at low energies, and this consistent set of string vacuums is known as landscape. With so many low energy descriptions, it may be difficult or impossible to find a vacuum to describe our world. Under this background, the swampland program \cite{Vafa:2005ui,Ooguri:2006in,ArkaniHamed:2006dz} was proposed. The swampland program aims to obtain consistent conditions for efficient field theory by requiring low energy efficiency field theory to always be coupled with quantum gravity. 

A convincing proposal for such a condition is the weak gravity conjecture (WGC) \cite{ArkaniHamed:2006dz}, which is a universal principle of string vacuums with uniform constraints. Various speculations have been proposed, but the efficiency field theory that appears as a low-energy description of quantum gravity theory must have a charge state greater than mass. If the above conjecture is not true, the extreme or near-extreme black hole will not be able to evaporate, because releasing the sub-extreme state will cause the remaining black hole to become a super-extreme, which violates the cosmic censorship. Various applications of the WGC and related quantum gravity conjectures have been discussed in some works \cite{Chen:2019qvr,Aalsma:2019ryi,Lee:2018spm,Lee:2019tst,Kats:2006xp,Loges:2019jzs,Montero:2019ekk,Cano:2019ycn,Cano:2019oma,Reall:2019sah,Heidenreich:2019zkl,Brahma:2019mdd,Gonzalo:2019gjp,Hamada:2018dde}. Among them, a method for judging WGC by the amount of change in entropy is mentioned \cite{Cheung:2018cwt}. This method points out: If $\Delta S>0$, then the single $U(1)$ form of the WGC follows immediately, the black hole systems display  WGC-like behavior \cite{Goon:2019faz}.

Recently, Goon and Penco proposed a universal thermodynamic extremality relation by the perturbative corrections of generic thermodynamic systems \cite{Goon:2019faz}
\begin{equation}
\frac{\partial M_{\mathrm{ext}}(\vec{\mathcal{Q}}, \epsilon)}{\partial \epsilon}=\lim _{M \rightarrow M_{\mathrm{ext}}(\vec{\mathcal{Q}}, \epsilon)}-T\left(\frac{\partial S(M, \vec{\mathcal{Q}}, \epsilon)}{\partial \epsilon}\right)_{M, \vec{\mathcal{Q}}}. \label{eq1}
\end{equation}
Where $\epsilon$ is the control parameter of the correction, $M_{ext}$, $T$, and $S$ are the extremality mass, temperature and entropy of the black hole respectively after the correction. And $Q$ denotes the extensive quantities of the black hole thermodynamics such as charge, angular momentum or other quantities. This relation was verified in the four-dimension RN-AdS black hole and extended to the rotating and massive gravity black holes by rescaling the cosmological constant as a perturbative correction \cite{Goon:2019faz,Wei:2020bgk,Chen:2020bvv}. 

In this paper, we will calculate the universal thermodynamic relations of black holes surrounded by quintessence. Nowadays, the nature of dark energy is driving the observed accelerated development of the universe, and this is undoubtedly one of the most exciting and challenging issues facing physicists and astronomers. It is one of the cores of current astronomical observations and proposals and is driving the way particle theorists try to understand the nature of the early and late universes. Astronomical observations indicate that the universe is accelerating, which implies the existence of a state of negative pressure. The cause of negative pressure can be twofold. One side is called quintessence \cite{Carroll:1991mt,Ratra:1987rm,Carroll:1998zi,Zlatev:1998tr,Peebles:1998qn}, with a possible explanation coming from the anti-gravity nature of dark energy. Quintessence can be described by an ordinary scalar field $\varphi$ that is minimally coupled with gravity, which has particular potentials that lead to late time inflation. We introduce a perturbative correction into the action and derive the extremality relations between the mass and pressure, entropy, charge, and quintessence factor, respectively.  Furthermore, we can make certain modifications to the black hole surrounded by the quintessence, such as considering the black hole under the nonlinear electrodynamics background \cite{AyonBeato:1998ub,AyonBeato:1999rg}, which is called regular Bardeen AdS black hole. A black hole has no singularity at the origin and has an event horizon. Bardeen first constructed a black hole solution with regular non-singular geometry and event horizon \cite{Bardeen}.

The rest of this paper is organized as follows. In the next section, the AdS black hole surrounded by the quintessence is given and its perturbative correction is discussed. We will use numerical analysis to calculate the extremality black hole mass $M_{ext}$ and entropy $S$ after small constant correction to the cosmological constant. In Sec. III, we will show the universal thermodynamic extremality relation to the regular Bardeen AdS black hole surrounded by quintessence field. In Sec. IV, we will summarize our research work.

\section{RN-AdS black holes surrounded by the quintessence}
We start with a simpler case and derive our hypothesis. We place the RN-AdS black hole in the quintessence considered as barotropic perfect fluid, and do not add other physical corrections. The bulk action for a RN-AdS black hole surrounded by quintessence dark energy in four dimensional is described as follows \cite{Kiselev,Ghaffarnejad:2018bsd}
\begin{equation}
\mathcal{S}=\frac{1}{16 \pi G} \int d^{4} x\left[\sqrt{-g}\left(R-2 \Lambda-F^{\mu \nu} F_{\mu \nu}\right)+\mathcal{L}_{q}\right].
\end{equation}
In this action, $R$ is the Ricci scalar, the cosmological constant is related to the AdS space radius $l$ by $\Lambda=-3 / l^{2}$ and $F_{\mu \nu}$ is the electromagnetic field tensor. The last term $\mathcal{L}_{q}$ in the action is the Lagrangian of quintessence as a barotropic perfect fluid, which is given by \cite{Minazzoli:2012md}
\begin{equation}
\mathcal{L}_{q}=-\rho_{q}\left[1+\omega_{q} \ln \left(\frac{\rho_{q}}{\rho_{0}}\right)\right],
\end{equation}
where $\rho_{q}$ is energy density, $\rho_{0}$ is the constant of integral, and the barotropic index $\omega_{q}$. Quintessence with the state equation given by the relation between the pressure $p_{q}$ and the energy density $\rho_{q}$, so that $p_{q}=\omega_{q} \rho_{q}$ at $\omega_{q}$ in the range of $-1<\omega_{q}<-1/3$ for the quintessence dark energy and $\omega_{q}<-1$ for the phantom dark energy, which causes the acceleration and is called barotropic index. The border case of $\omega_{q}=-1$ of the extraordinary quintessence covers the cosmological constant term. 

The metric of the charged RN-AdS black hole surrounded by quintessence is
\begin{align}
d s^{2} &=-f(r) d t^{2}+\frac{1}{f(r)} d r^{2}+r^{2}(r)\left(d \theta^{2}+\sin ^{2} \theta d \varphi^{2}\right), \nonumber \\
f(r) &=1-\frac{2 M}{r}+\frac{Q^{2}}{r^{2}}+\frac{r^{2}}{l^{2}}-\frac{a}{r^{3 \omega_{q}+1}}.
\end{align}
Where $M$ and $Q$ are the mass and electric charge of the black hole, respectively, and $a$ is the
 factor related to the density of quintessence as \cite{AzregAinou:2012hy,Azreg-Ainou:2014lua}
\begin{equation}
\rho_{q}=-\frac{a}{2} \frac{3 \omega_{q}}{r^{3}\left(\omega_{q}+1\right)}.
\end{equation}
We use the black hole entropy $S = \pi r _ +$ and combine the black hole horizon equation, the mass $M$ and temperature $T$ of this black hole can be expressed as
\begin{align}
T&=\frac{3}{4} a \pi ^{\frac{3 \omega_{q} }{2}} \omega_{q} S^{-\frac{3 \omega_{q} }{2}-1}+\frac{3 \sqrt{S}}{4 \pi ^{3/2} l^2}-\frac{\sqrt{\pi } Q^2}{4S^{3/2}}+\frac{1}{4 \sqrt{\pi } \sqrt{S}},\\
M&=-\frac{1}{2} a \pi ^{\frac{3 \omega_{q} }{2}} S^{-\frac{3 \omega_{q} }{2}}+\frac{S^{3/2}}{2 \pi ^{3/2} l^2}+\frac{\sqrt{\pi } Q^2}{2 \sqrt{S}}+\frac{\sqrt{S}}{2 \sqrt{\pi }}.
\end{align}

\begin{figure}[H]
	\centering
	\subfigure[unmodified $M$ as a function of charge $Q$]{
		\includegraphics[width=0.45\linewidth]{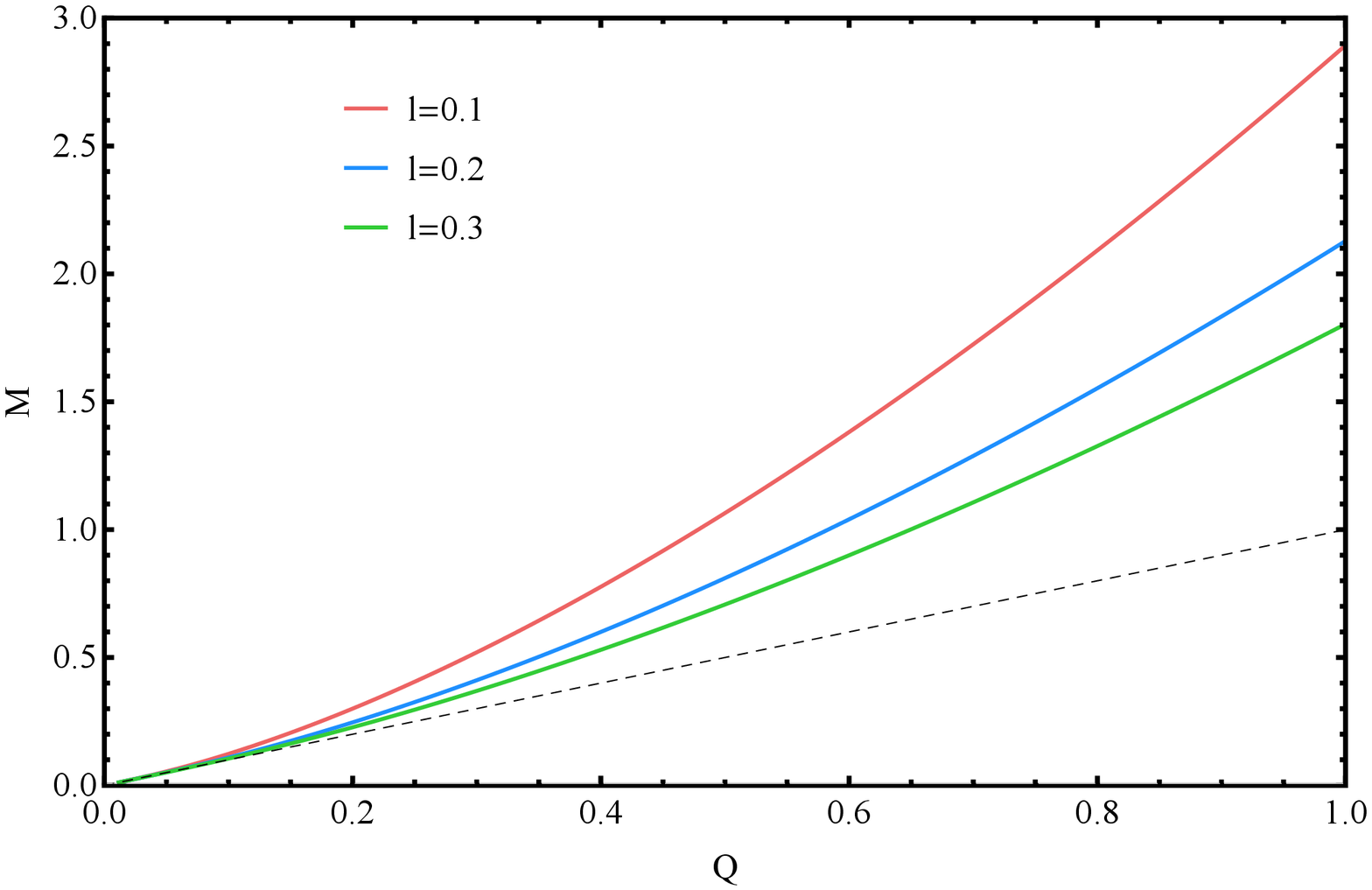} 
	}
	\subfigure[unmodified $S$ as a function of charge $Q$]{
		\includegraphics[width=0.45\linewidth]{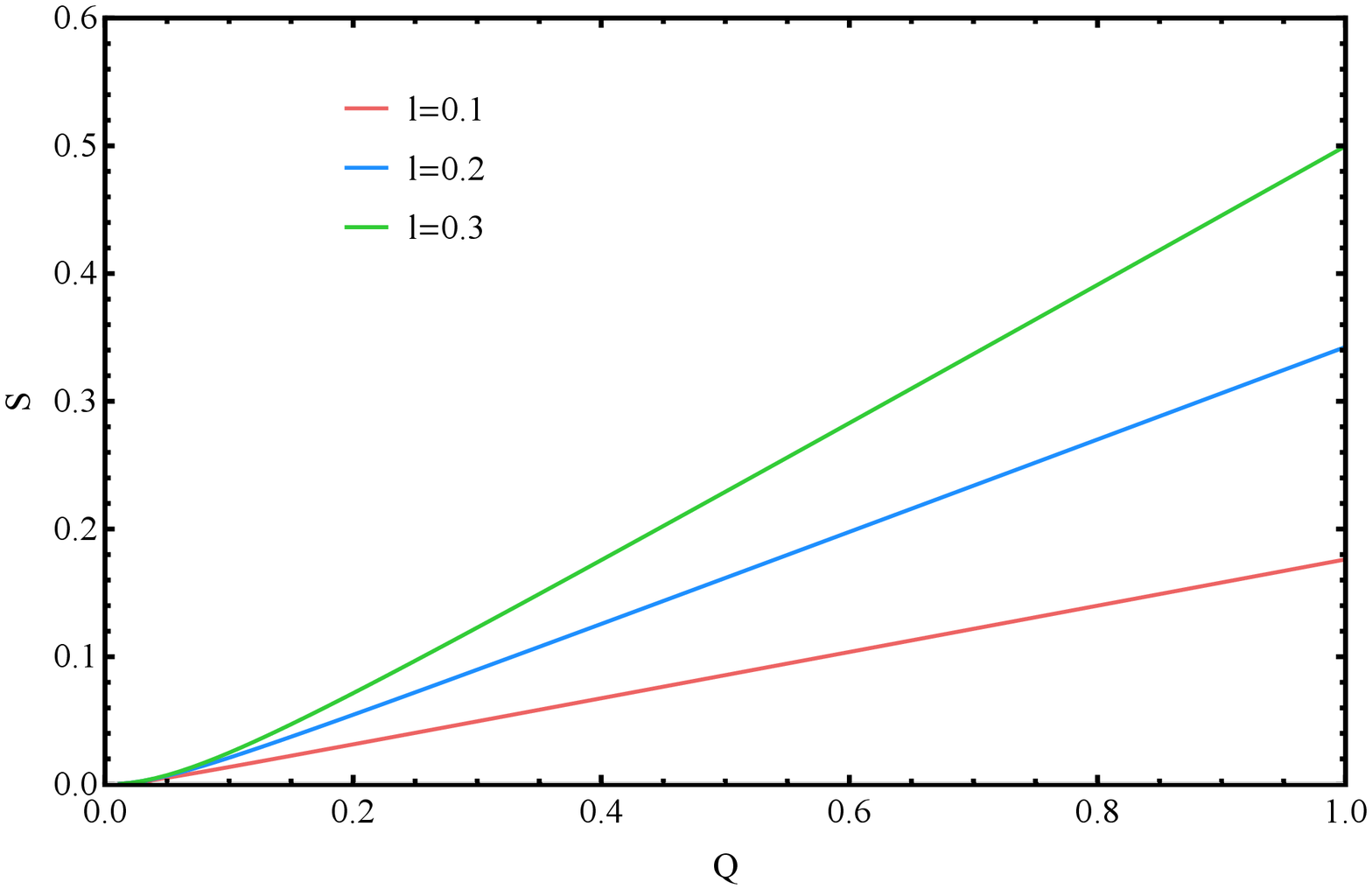}
	}
	\caption{The parameters of the black hole vary with the black hole charge $Q$. We fixed density of quintessence $\rho_{q}$ as 1.0, $\omega_{q}$ as $-2/3$ and AdS space radius $l$ as $\text{0.1, 0.2, 0.3}$ in order. The solid line in the figure is the curve of the black hole mass varies with the black hole charge, and the black hole entropy varies with the black hole charge. And the dashed line in the figure is the critical curve where the ratio of the black hole mass to the charge is one.}
	\label{M-S-uncorrected}
\end{figure}

Moreover, we can write those physical quantity $\eta=-\frac{1}{2} \pi ^{\frac{3 \omega_{q}}{2}} S^{-\frac{3 \omega_{q}}{2}}$  conjugate  to the parameter $a$, electric potential $\Phi=\sqrt{\pi } Q/\sqrt{S}$ and thermodynamic volume $V=4 S^{3/2}/3 \sqrt{\pi }$. For comparing with the modified mass $M$ and entropy $S$, we fixed some parameters here to draw a graph of the unmodified mass $M$ and entropy $S$ as a function of the charge $Q$ when the black hole near-extreme. From the picture on the left of FIG. (\ref{M-S-uncorrected}), we draw the critical state where the mass-to-charge ratio is equal to one, which is represented by a dashed line. We can find that the ratio of the unmodified black hole mass to the charge amount is greater than one. From the picture on the right of FIG. (\ref{M-S-uncorrected}), we can find that the unmodified black hole entropy increases as the black hole charge increases.

\begin{figure}[H]
	\centering
	\subfigure[modified $M$ as a function of charge $Q$ compares with unmodified $M$ as a function of charge $Q$. ]{
		\includegraphics[width=0.45\linewidth]{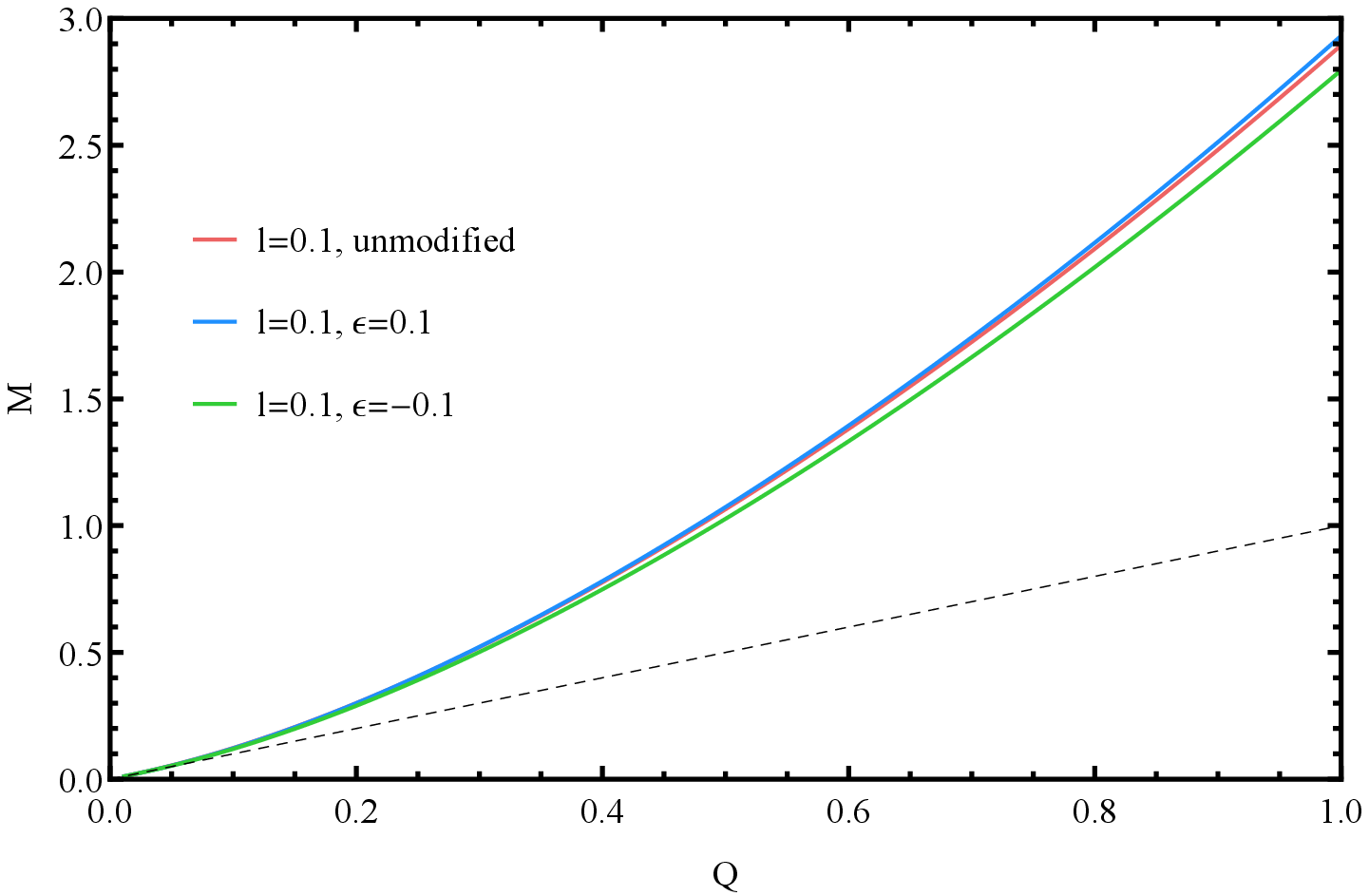} 
	}
	\subfigure[modified $S$ as a function of charge $Q$ compares with unmodified $S$ as a function of charge $Q$. ]{
		\includegraphics[width=0.45\linewidth]{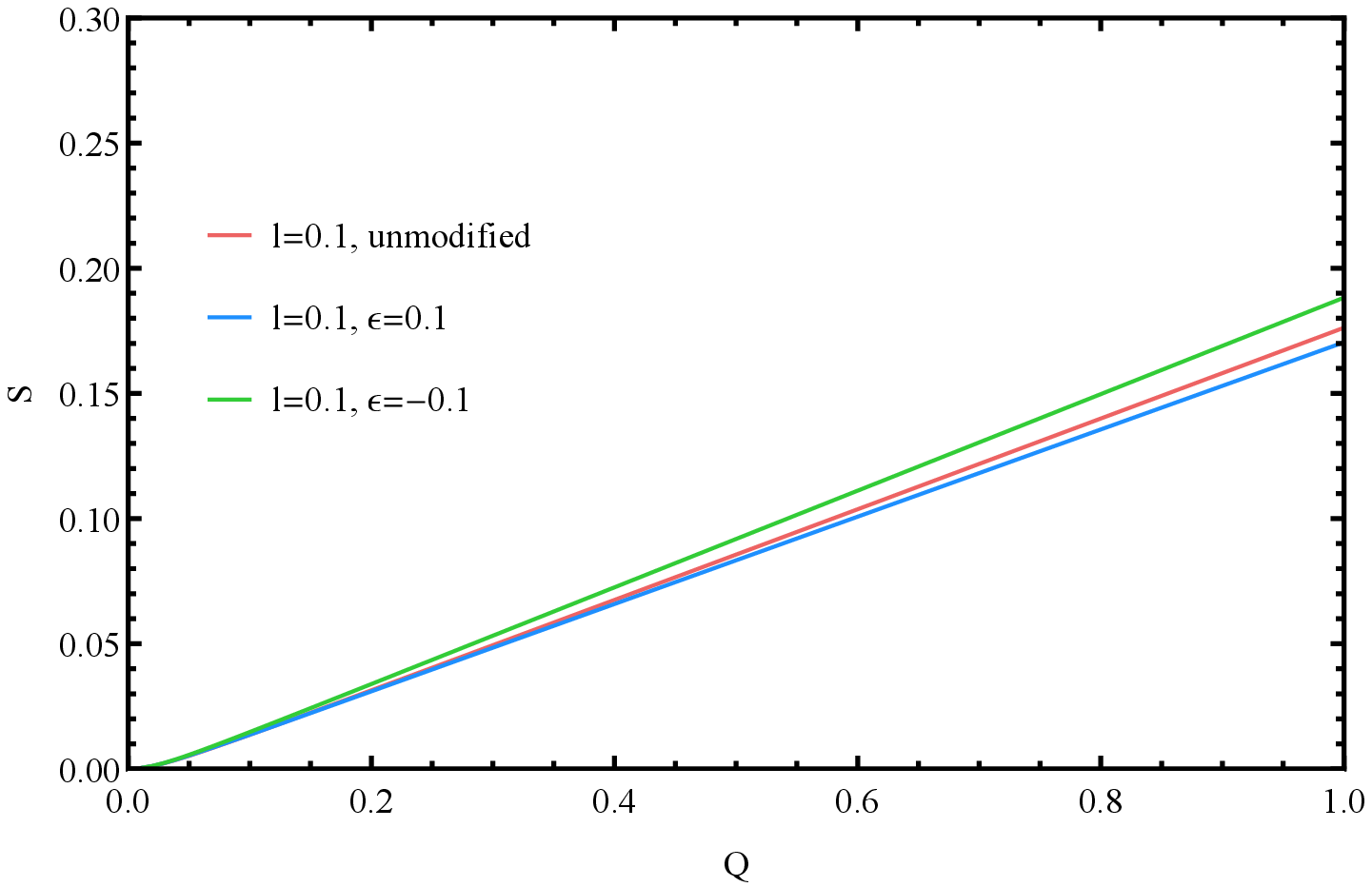}
	}
	\caption{The parameters of the black hole vary with the black hole charge $Q$. We fixed the normalization factor $a$ as 1.0, AdS space radius $l$ as 0.1 and $\omega_{q}$ as $-2/3$. Besides, we fixed the small constant $\epsilon$ as -0.1, 0.0, and 0.1, which are represented by green, red and blue solid lines respectively. And, the dashed line represents the critical state which means the mass-to-charge ratio is equal to one.}
	\label{M-S-corrected}
\end{figure}

And then, We make a small correction to the cosmological constant term in the action, ones can obtain
\begin{equation}
\mathcal{S}=\frac{1}{16 \pi G} \int d^{4} x\left\lbrace \sqrt{-g}\left[R-2(1+\epsilon) \Lambda-F^{\mu \nu} F_{\mu \nu}\right]+\mathcal{L}_{q}\right\rbrace,
\end{equation}
and the modified mass $M$ and temperature $T$ 
\begin{align}
T&=\frac{3}{4} a \pi ^{\frac{3 \omega_{q}}{2}} \omega_{q}  S^{-\frac{3 \omega_{q} }{2}-1}+\frac{3 \sqrt{S} (\epsilon +1)}{4 \pi ^{3/2} l^2}-\frac{\sqrt{\pi } Q^2}{4 S^{3/2}}+\frac{1}{4 \sqrt{\pi } \sqrt{S}},\label{eqTe}\\
M&=-\frac{1}{2} a \pi ^{\frac{3 \omega_{q}}{2}} S^{-\frac{3 \omega_{q}}{2}}+\frac{S^{3/2} (\epsilon +1)}{2 \pi ^{3/2} l^2}+\frac{\sqrt{\pi } Q^2}{2 \sqrt{S}}+\frac{\sqrt{S}}{2 \sqrt{\pi }}.\label{eqMe}
\end{align}
With these expressions, we can do a numerical analysis of the black hole mass and black hole entropy corrected with a small constant $\epsilon$. From the picture on the left of FIG. (\ref{M-S-corrected}), we find the mass of black hole has been some small perturbations if we compare them unmodified black hole mass. When $\epsilon=0.1$, the mass of the black hole increases, and when $\epsilon=-0.1$, the mass of the black hole decreases. From the picture on the right of FIG. (\ref{M-S-corrected}), we can obtain that the black hole entropy also has been some small perturbation similar to the black hole mass. However, unlike the black hole mass perturbation, the condition of the black hole entropy increases and decreases in the opposite. When $\epsilon=0.1$, the entropy of the black hole decreases, and when $\epsilon=-0.1$, the entropy of the black hole increases. From here, we can preliminarily see the clue of WGC. When our constant correction $\epsilon$ to be negative, the mass of the black hole can be lower, the mass-charge ratio of the black hole gradually decreases and approaches to one. The change of the entropy $\Delta S>0$ of the black hole is consistent with the criterion pointed out in \cite{Cheung:2018cwt}.

From Eq. (\ref{eqMe}), we can obtain the small correction $\epsilon$
\begin{align}
\epsilon=a l^2 \pi ^{\frac{3 \omega_{q}+3 }{2}} S^{-\frac{3 \omega_{q}+3 }{2}}+\frac{2 \pi ^{3/2} l^2 M}{S^{3/2}}-\frac{\pi ^2 l^2 Q^2}{S^2}-\frac{\pi  l^2}{S}-1. \label{epsilon}
\end{align}
Taking the derivative of it with respect to S, one can get
\begin{align}
\left(\frac{\partial \epsilon}{\partial S}\right)_{M, Q,l,a}=-\frac{3}{2}a l^2 \pi ^{\frac{3 \omega_{q}+3}{2}} \omega_{q}  S^{-\frac{3 \omega_{q}-5}{2}}+\frac{\pi ^2 l^2 Q^2}{2 S^3}-\frac{\pi  l^2}{2 S^2}-\frac{3 (\epsilon+1)}{2 S} \label{eps}.
\end{align}
Combining equations (\ref{eqTe}) and (\ref{eps}), we can obtain
\begin{equation}
-T\left(\frac{\partial S}{\partial \epsilon}\right)_{M,Q, l,a}=\frac{S^{3/2}}{2 \pi ^{3/2} l^2}.\label{eq27}
\end{equation}

Let's consider $\left(\frac{\partial M_{e x t}}{\partial \epsilon}\right)_{a,l,Q}$ now. The new extremality can be obtained as $T=0$, it's very difficult to solve for entropy directly. We can express extremal mass as $M_{\mathrm{ext}}\left(a,l,Q,S(a,l,Q,\epsilon),\epsilon\right)$, then use the chain rule of derivation
\begin{equation}
\left.\frac{\partial M_{\mathrm{ext}}}{\partial \epsilon}\right|_{a, l,Q}=\left.\left.\frac{\partial M_{\mathrm{ext}}}{\partial S}\right|_{a, l, Q,\epsilon} \frac{\partial S}{\partial \epsilon}\right|_{a, l, Q,ext}+\left.\frac{\partial M_{\mathrm{ext}}}{\partial \epsilon}\right|_{a, l, Q,S},
\end{equation}
where $\left.\frac{\partial S}{\partial \epsilon}\right|_{a, l, Q,ext}$ can be obtained by extremality condition. And $\left.\frac{\partial M_{\mathrm{ext}}}{\partial S}\right|_{a, l, Q,\epsilon}$ and $\left.\frac{\partial M_{\mathrm{ext}}}{\partial \epsilon}\right|_{a, l, Q,S}$ are partial derivatives of Eq. (\ref{eqMe}). When Eq. (\ref{eqTe}) equals to zero, we have
\begin{align}
\left.\frac{\partial \epsilon}{\partial S}\right|_{a, l, Q,ext}=\frac{3}{2} a l^2 (\omega_{q}+1) \pi ^{\frac{3 \omega_{q}}{2}+\frac{3}{2}} \omega_{q}  S^{-\frac{3 \omega_{q} }{2}-\frac{5}{2}}-\frac{2 \pi ^2 l^2 Q^2}{3 S^3}+\frac{\pi  l^2}{3 S^2}.
\end{align}
Combining these results, we can obtain
\begin{equation}
\left.\frac{\partial M_{\mathrm{ext}}}{\partial \epsilon}\right|_{a, l,Q}=\frac{S^{3/2}}{2 \pi ^{3/2} l^2}=-T\left(\frac{\partial S}{\partial \epsilon}\right)_{M,Q, l,a}.
\end{equation}
Similarly, we continue to investigate the extremality relation between the mass and charge. From Eq. (\ref{epsilon}), the differential of  $\epsilon$ to $Q$ takes the form
\begin{align}
\left(\frac{\partial \epsilon}{\partial Q} \right)_{ M,S,l,a}&=-\frac{2 \pi  l^2 Q}{S^2},.
\end{align}
Multiplying by the electric potential $\Phi=\sqrt{\pi } Q/\sqrt{S}$ and taking the extremality, we have
\begin{align}
-\Phi\left(\frac{\partial Q}{\partial \epsilon}\right)_{M,S,l,a}&=\frac{S^{3/2}}{2 \pi ^{3/2} l^2}=\left.\frac{\partial M_{\mathrm{ext}}}{\partial \epsilon}\right|_{a, l,Q}.
\end{align}

Next, we would like to regard the cosmological constant as a variable related to pressure, and investigate the extremality relation between the mass and pressure.  Bring the relation $P=3/8\pi l^{2}$ to Eq. (\ref{epsilon}), the differential of  $P$ to $\epsilon$  takes the form
\begin{align}
\left(\frac{\partial P}{\partial \epsilon}\right)_{M,S,Q,a}&=\frac{3 \left(-a \pi ^{\frac{3 \omega_{q}+1}{2}} S^{\frac{1-3 \omega_{q}}{2}}-2 \sqrt{\pi } M \sqrt{S}+\pi  Q^2+S\right)}{8 S^2 (\epsilon +1)^2}.
\end{align}
Multiplying by the thermodynamic volume $V=4 S^{3/2}/3 \sqrt{\pi }$ and taking the extremality, we have
\begin{align}
-V\left(\frac{\partial P}{\partial \epsilon}\right)_{M,S,Q,a}=\frac{S^{3/2}}{2 \pi ^{3/2} l^2}=\left.\frac{\partial M_{\mathrm{ext}}}{\partial \epsilon}\right|_{a, l,Q}.
\end{align}

Last, we can consider the extremality relation between the mass and conjugate quantity $\eta$. From Eq. (\ref{epsilon}), the differential of quintessence parameter $a$ to  $\epsilon$ takes the form
\begin{align}
\left(\frac{\partial a}{\partial \epsilon}\right)_{ M,S,Q,l}&=\frac{\pi ^{-\frac{3 \omega }{2}-\frac{3}{2}} S^{\frac{3 \omega }{2}+\frac{3}{2}}}{l^2}.\label{eq23}
\end{align}
Multiplying by $\eta=-\frac{1}{2} \pi ^{\frac{3 \omega_{q}}{2}} S^{-\frac{3 \omega_{q}}{2}}$ and taking the extremality, we have
\begin{align}
-\eta\left(\frac{\partial a}{\partial \epsilon}\right)_{M,S,Q,l}=\frac{S^{3/2}}{2 \pi ^{3/2} l^2}=\left.\frac{\partial M_{\mathrm{ext}}}{\partial \epsilon}\right|_{a, l,Q}.
\end{align}

\begin{figure}[H]
	\centering
	\subfigure[modified $M$ as a function of parameter $\epsilon$]{
		\includegraphics[width=0.45\linewidth]{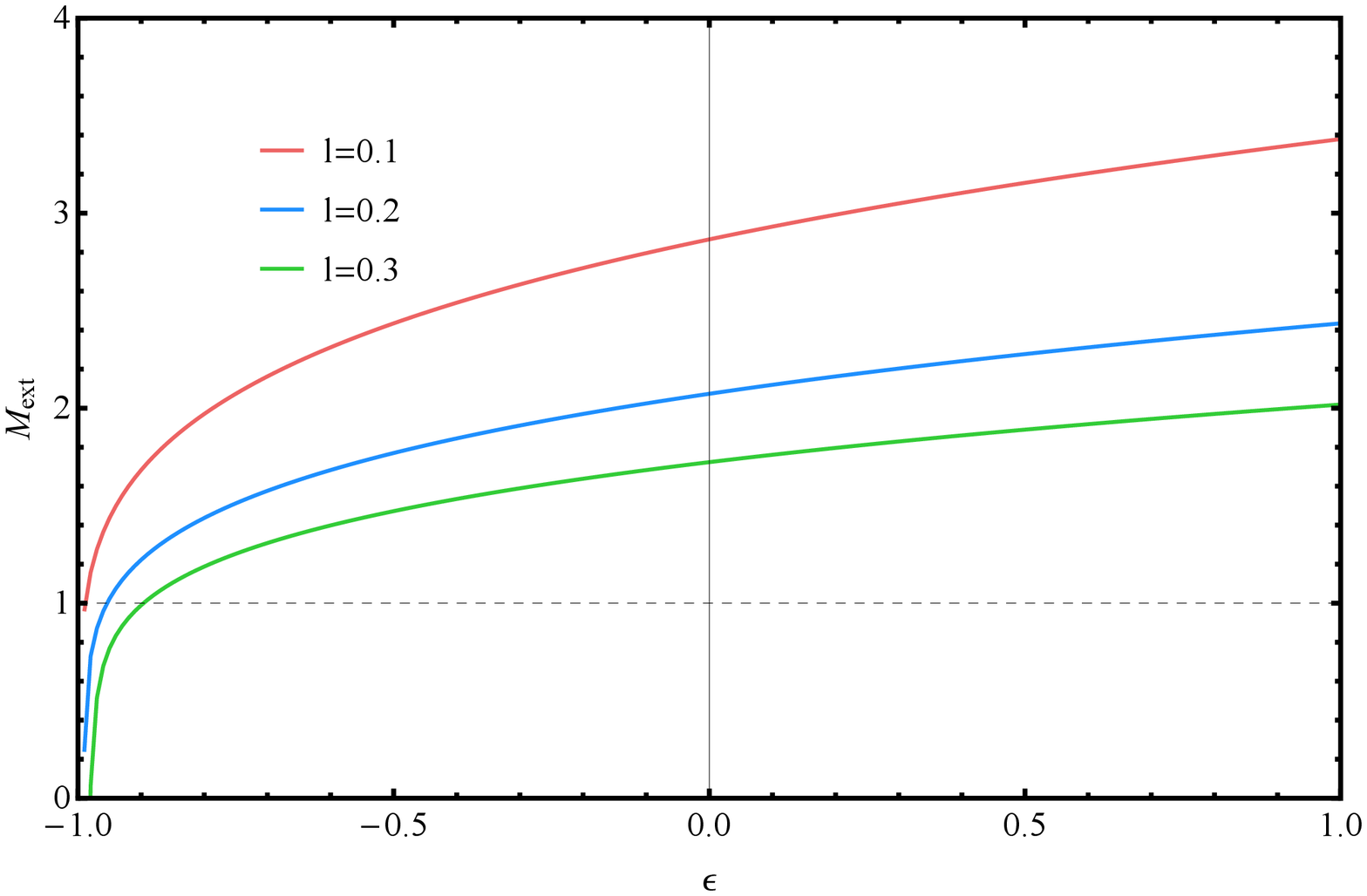} 
	}
	\subfigure[modified $S$ as a function of parameter $\epsilon$]{
		\includegraphics[width=0.45\linewidth]{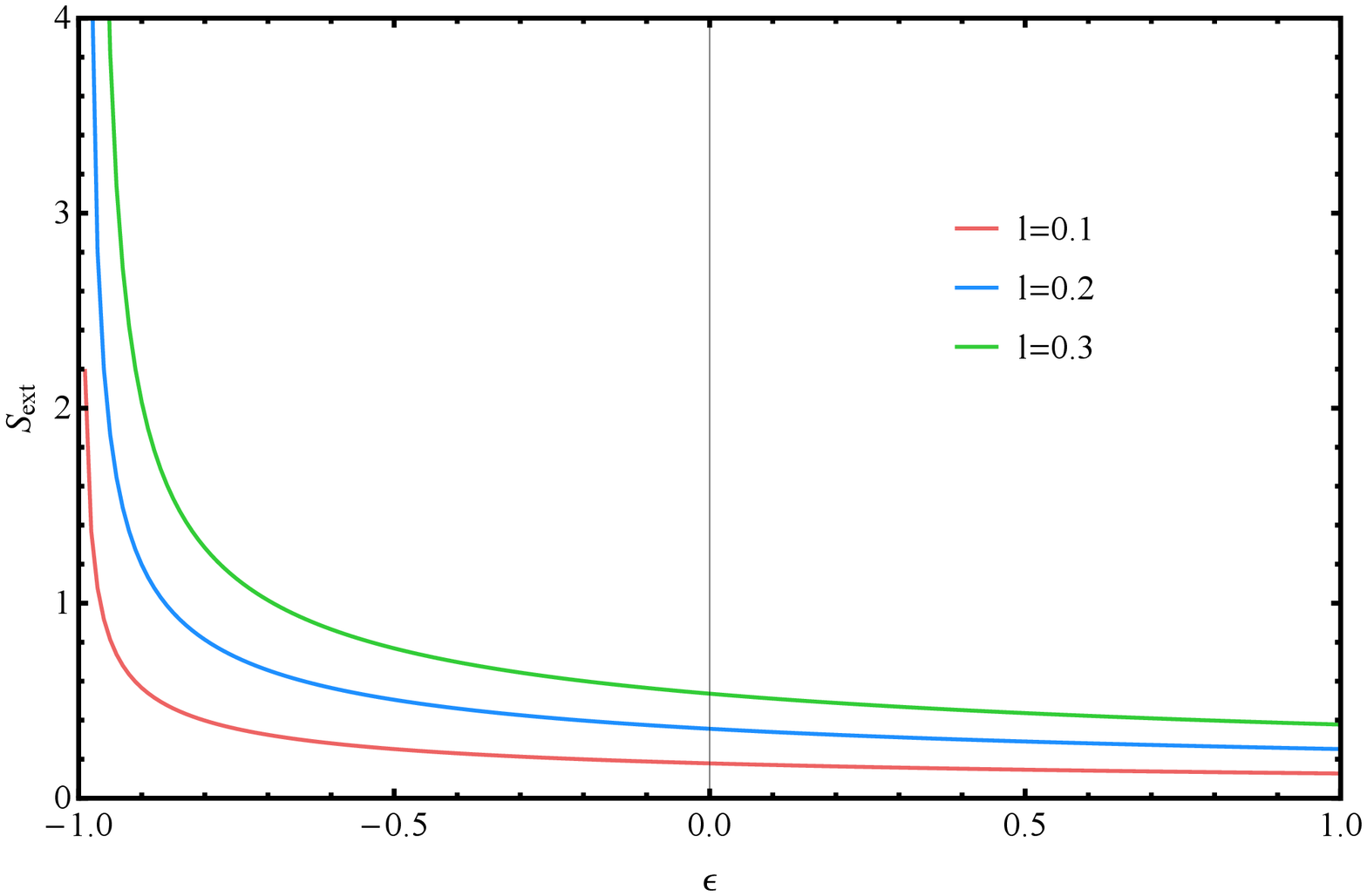}
	}
	\caption{The parameters of the black hole vary with the small correction parameter $\epsilon$. We fixed normalization factor $a$ as 1.0, $\omega_{q}$ as $-2/3$, black hole charge $Q$ as 1.0, and AdS space radius $l$ as $\text{0.1, 0.2, 0.3}$ in order. The color solid lines in the figure is the curve of black hole mass and entropy varies with small correction parameter $\epsilon$. The dashed line in the figure is the critical curve with the black hole mass equaling to charge. The intersection of the black vertical line and the color solid lines at $\epsilon=0$ in the figure indicates the uncorrected mass and entropy of the black hole when $Q=1.0$.}
	\label{M-S-epsilon}
\end{figure}

Therefore, we confirm the Goon-Penco extremality relation for the charged RN-AdS black hole surrounded by quintessence under the constant correction of the action. In order to show how the parameter $\epsilon$ modifies the black hole mass and black hole entropy, we perform numerical analysis on corrected mass and entropy. In such a case, when $\epsilon=0$, the system degenerates to the situation without correction. From the figure on the left of FIG. (\ref{M-S-epsilon}), we find that the mass of the black hole increases when the correction constant $\epsilon$ is positive, and decreases when the correction constant $\epsilon$ is negative. In addition, the larger the radius $l$ of the AdS space, the easier it is to correct the black hole mass-charge ratio $M/Q$ to less than one, which is what we expect. From the figure on the right of FIG. (\ref{M-S-epsilon}), we can obtain that the entropy of the black hole decreases when the correction constant $\epsilon$ is positive, and increases when the correction constant $\epsilon$ is negative. In other word, when the black hole entropy increases, the correction constant is negative, the black hole mass and the mass-charge ratio decreases, hence this black hole displays WGC-like behavior.

\section{Regular Bardeen AdS black hole surrounded by quintessence field}
Based on the Sec. II, we consider the universal thermodynamic extremality relations under a nonlinear electrodynamic source coupled to AdS spacetime with quintessence field. This type of black hole can be called as regular Bardeen AdS black hole.

The action for regular Bardeen AdS black hole surrounded by quintessence field is given by \cite{AyonBeato:1998ub,AyonBeato:1999rg}
\begin{equation}
S=\int d^{4} x \sqrt{-g}\left[\frac{1}{16 \pi} R-\frac{1}{8 \pi}\Lambda-\frac{1}{4 \pi} \mathcal{L}(\mathcal{F})\right].
\end{equation}
Where $R$ is the Ricci scalar, the cosmological constant is related to the AdS space radius $l$ by $\Lambda=-3 / l^{2}$, and $\mathcal{L}(\mathcal{F})$ represents the Lagrangian for a nonlinear electrodynamic source 
\begin{equation}
\mathcal{L}(\mathcal{F})=\frac{3 M}{\beta^{3}}\frac{\sqrt{4 \beta^{2} \mathcal{F}}}{1+\sqrt{4 \beta^{2} \mathcal{F}}}.
\end{equation}
Where $\mathcal{F}=F_{\mu \nu} F^{\mu \nu}$, $\beta$ is the magnetic monopole charge of a self gravitating magnetic field described by nonlinear electromagnetic source. The solution for this action is given by
\begin{align}
d s^{2}&=-f(r) d t^{2}+\frac{1}{f(r)} d r^{2}+r^{2}(r)\left(d \theta^{2}+\sin ^{2} \theta d \varphi^{2}\right),\nonumber\\
f(r)&=1-\frac{2 M r^{2}}{\left(\beta^{2}+r^{2}\right)^{3 / 2}}+\frac{r^{2}}{l^{2}}-\frac{a}{r^{3 \omega_{q}+1}}.
\end{align}
The following work is the same as in the Section II. We make a small correction to the cosmological constant term in the action
\begin{equation}
\mathcal{S}=\int d^{4} x \sqrt{-g}\left[\frac{1}{16 \pi} R-\frac{(1+\epsilon)}{8 \pi}\Lambda-\frac{1}{4 \pi} \mathcal{L}(\mathcal{F})\right].
\end{equation}
We can obtain the modified mass $M$ and temperature $T$, which are described by entropy $S$, magnetic monopole charge $\beta$ and normalization factor $a$
\begin{align}
T&=\frac{3}{4} a \pi ^{\frac{3 \omega_{q} }{2}} \sqrt{\pi  \beta ^2+S} S^{-\frac{3 \omega_{q}}{2}-\frac{5}{2}} \left[\pi  \beta ^2 (\omega_{q} +1)+S \omega_{q} \right]\nonumber\\
&+\frac{3 (\epsilon +1) \sqrt{\pi  \beta ^2+S}}{4 \pi ^{3/2} l^2}+\frac{\left(S-2 \pi  \beta ^2\right) \sqrt{\pi\beta ^2+S}}{4\pi^{1/2} S^2},\label{eqT}\\
M&=\frac{\left(\pi  \beta ^2+S\right)^{3/2} \left(S^{\frac{3 \omega_{q}+1}{2}}-a \pi ^{\frac{3 \omega_{q}+1}{2}}\right)}{2 \pi^{1/2}S^{\frac{3\omega_{q} +3}{2}}}+\frac{(\epsilon +1) \left(\pi\beta ^2+S\right)^{3/2}}{2 \pi^{3/2}l^2}.\label{eqM}
\end{align}

In order to calculate the left side of Eq.(\ref{eq1}), we need to insert the entropy at the extremality into the mass (\ref{eqM}) to obtain the differential relationship. It's very difficult to do that, however, we can use the chain rule of derivation 
\begin{equation}
\left.\frac{\partial M_{\mathrm{ext}}}{\partial \epsilon}\right|_{a,l,\beta}=\left.\left.\frac{\partial M_{\mathrm{ext}}}{\partial S}\right|_{a,l,\beta, \epsilon} \frac{\partial S}{\partial \epsilon}\right|_{a,l,\beta, ext}+\left.\frac{\partial M_{\mathrm{ext}}}{\partial \epsilon}\right|_{a,l,\beta, S},
\end{equation}
where $\left.\frac{\partial S}{\partial \epsilon}\right|_{a, l, \beta,ext}$ can be obtained by extremality condition. And $\left.\frac{\partial M_{\mathrm{ext}}}{\partial S}\right|_{a, l, \beta,\epsilon}$ and $\left.\frac{\partial M_{\mathrm{ext}}}{\partial \epsilon}\right|_{a, l, \beta,S}$ are partial derivatives of Eq. (\ref{eqT}). When Eq. (\ref{eqT}) equals to zero, we have
\begin{align}
\left.\frac{\partial \epsilon}{\partial S}\right|_{a, l,\beta, ext}&=\frac{\pi  l^2 }{6 S^{\frac{3 \omega_{q} +7}{2}}}\left[3 a \beta ^2 \pi ^{\frac{3 \omega_{q}+3}{2}} \left(3 \omega_{q} ^2+8\omega_{q} +5\right)+9 a S \pi ^{\frac{3 \omega_{q}+1 }{2}} \omega_{q}  (\omega_{q} +1)-8 \pi  \beta ^2 S^{\frac{3 \omega_{q} +1}{2}}+2 S^{\frac{3 \omega_{q} +3}{2}}\right].
\end{align}
Therefore, we can get
\begin{equation}
\left.\frac{\partial M_{\mathrm{ext}}}{\partial \epsilon}\right|_{a, l,\beta}=\frac{\left(\pi \beta ^2+S\right)^{3/2}}{2 \pi^{3/2}l^2}.
\end{equation}

The next process is similar to the black hole which we derived in Sec. II. We continue to calculate the extremality relation between the mass and thermodynamic quantities such as entropy etc. From Eq. (\ref{eqM}), we can obtain the small correction $\epsilon$ as
\begin{align}
\epsilon=a l^2 \pi ^{\frac{3\omega_{q} +3}{2}} S^{-\frac{3\omega_{q} +3}{2} }+2\pi^{3/2} M l^2\left(\pi\beta ^2+S\right)^{-3/2}-\pi l^2 S^{-1} -1.\label{ep1}
\end{align}
Taking the derivative of it with respect to $S$, one can get 
\begin{equation}
\left(\frac{\partial \epsilon}{\partial S}\right)_{M,a,l, \beta}=-\frac{3}{2}(\omega_{q} +1)a l^2 \pi ^{\frac{3\omega_{q} +3}{2}} S^{-\frac{3\omega_{q} +5}{2} }-3\pi^{3/2} M l^2\left(\pi\beta ^2+S\right)^{-5/2}+\pi l^2 S^{-2}.
\end{equation}
And, taking the derivative of it with respect to $a$, one can obtain 
\begin{equation}
\left(\frac{\partial \epsilon}{\partial a}\right)_{M,S, \beta,l}=l^2 \pi ^{\frac{3\omega_{q} +3}{2}} S^{-\frac{3\omega +3}{2}}.
\end{equation}
Then, we using the relations $\eta=-\frac{1}{2} \pi ^{\frac{3\omega_{q} }{2}} \left(\pi \beta ^2+S\right)^{3/2} S^{-\frac{3\omega_{q} +3}{2}}$ and the temperature $T$ to obtain the following relations
\begin{equation}
\left.\frac{\partial M_{\mathrm{ext}}}{\partial \epsilon}\right|_{a, l,\beta}=\frac{\left(\pi \beta ^2+S\right)^{3/2}}{2 \pi^{3/2}l^2}=-T\left(\frac{\partial S}{\partial \epsilon}\right)_{M, a,l,\beta}=-\eta\left(\frac{\partial a}{\partial \epsilon}\right)_{M,S,l,\beta}. 
\end{equation}

Consider the effective potential $\mathcal{B}$ which conjugates to $\beta^{2}$ with small correction parameter $\epsilon$, which can be obtained from the first law of thermodynamics as
\begin{equation}
\mathcal{B}=\frac{3 \sqrt{\pi\beta^{2}+S} }{4\pi^{1/2} l^2}\left(\pi  l^2+S \epsilon +S-a l^2 \pi ^{\frac{3\omega_{q} }{2}+\frac{1}{2}}S^{-\frac{3}{2}\omega_{q}-\frac{1}{2}}\right).\label{eqA}
\end{equation}
Combining the Eq. (\ref{ep1}) and Eq. (\ref{eqA}), we can find that
\begin{align}
-\mathcal{B}\left(\frac{\partial \beta^{2}}{\partial \epsilon}\right)_{M,S,a,l}&=\frac{\left(\pi \beta ^2+S\right)^{3/2}}{2 \pi^{3/2}l^2}=\left.\frac{\partial M_{\mathrm{ext}}}{\partial \epsilon}\right|_{a, l,\beta}.
\end{align}
Thus, we find a new form of the extremality relation (\ref{eq1}). From above, we confirm the Goon-Penco relation for the regular Bardeen AdS black hole surrounded by quintessence field.
\section{Conclusion}
In this paper, we investigated the thermodynamic relation when a constant correction to the action is included in the charged  black holes surrounded by quintessence. We confirmed the extremality relations between the mass and entropy, charge, pressure which derived by \cite{Goon:2019faz,Wei:2020bgk,Chen:2020bvv} first.
We proposed a new effective parameter $\eta$ and potential $\mathcal{B}$ for the charged  black holes surrounded by quintessence, and confirmed it for both the two black holes. In four dimensional RN-AdS black hole, we obtain extremality relations between black hole mass and the normalization factor $a$ related to the density of quintessence 
\begin{align}
\left(\frac{\partial M_{ext}}{\partial \epsilon}\right)_{a,l,Q}=\lim _{M \rightarrow M_{ext}}-\eta\left(\frac{\partial a}{\partial \epsilon}\right)_{M,S,l,Q}.
\end{align}
And, in four dimensional regular Bardeen black hole, we obtain the extremality relation between black hole mass and the magnetic monopole charge $\beta$ 
\begin{align}
\left(\frac{\partial M_{ext}}{\partial \epsilon}\right)_{a,l,\beta}=\lim _{M \rightarrow M_{e x t}}-\mathcal{B}\left(\frac{\partial \beta^{2}}{\partial \epsilon}\right)_{M,S,a,l}.
\end{align}
In Ref. \cite{Wei:2020bgk}, the researches assumed a general formula of the extremality relation existed in black holes. Our work gives a partial verification to this formula.

By using this new method of proving the WGC, we make small constant correction $\epsilon$ to the cosmological constant $\Lambda$ term in the action of the black hole. This causes the perturbation on the mass and entropy of the black hole. When the correction constant $\epsilon$ is less than zero, the extremality mass of the black hole can be reduced, and the entropy of the black hole can increase, which has been shown in FIG. (\ref{M-S-epsilon}). In this case, the mass-charge ratio $M/Q$ of the black hole will gradually decreases and approaches to one, and $\Delta S>0$ satisfies the criterion in the article \cite{Cheung:2018cwt}. It is consistent with our expected, black hole systems display WGC-like behavior.  In addition, we can find it from FIG. (\ref{M-S-epsilon}) that the larger the radius $l$ of the AdS spacetime, the easier it is for the black hole to be corrected to the $M<Q$ state by a small constant correction parameter.

\begin{acknowledgments}
We are grateful to Deyou Chen, Yuzhou Tao and Peng Wang for useful discussions. This work is supported by NSFC (Grant No. 11947408).  
\end{acknowledgments}

\end{document}